\documentclass[12pt]{article}
\usepackage{amsmath}
\usepackage{longtable}
\usepackage{graphicx,psfrag,epsf}
\usepackage{enumerate}
\usepackage{natbib}
\usepackage{setspace}
\usepackage{authblk}
\usepackage{hyperref}
\usepackage{url} 
\newcommand{\blind}{0}
\usepackage{amssymb}
\usepackage{xr}
\newcommand{\bfy}{\mathbf{y}}
\newcommand{\Dirichlet}{{\textnormal{Dirichlet}}}
\newcommand{\gammadist}{{\textnormal{Gamma}}}

\newcommand{\bfone}{\mathbf{1}}
\newcommand{\bfmu}{\pmb{\mu}}
\newcommand{\bfeta}{\pmb{\eta}}

\newcommand{\bfX}{\pmb{X}}
\newcommand{\bfx}{\pmb{x}}

\newcommand{\bfz}{\pmb{z}}
\newcommand{\bfY}{\pmb{Y}}
\newcommand{\bfA}{\pmb{A}}
\newcommand{\bfB}{\pmb{B}}
\newcommand{\bfbeta}{\pmb{\beta}}
\newcommand{\bftheta}{\pmb{\theta}}
\newcommand{\bfgamma}{\pmb{\gamma}}
\newcommand{\airbnb}{Airbnb }
\newcommand{\airbnbp}{Airbnb}

\makeatletter
\newcommand*{\addFileDependency}[1]{
  \typeout{(#1)}
  \@addtofilelist{#1}
  \IfFileExists{#1}{}{\typeout{No file #1.}}
}
\makeatother

\newcommand{\alr}{{\textnormal{alr}}}

\newcommand{\bias}{{\text{bias}}}
\newcommand{\RMSE}{{\text{RMSE}}}

\newcommand{\FMAE}{{\text{FMAE}}}

\newcommand{\FRMSE}{{\text{FRMSE}}}

\newcommand{\CIL}{{\text{CIL}}}
\newcommand{\elpd}{{\text{ELPD}}}

\newcommand{\yearly}{{\text{year}}}
\newcommand{\week}{{\text{week}}}
\newcommand{\season}{{\text{season}}}
\newcommand{\LFO}{{\text{LFO}}}
\newcommand{\IN}{{\text{\% IN}}}

\title{A Bayesian Dirichlet Auto-Regressive Moving Average Model for Compositional Time Series}

\def\spacingset#1{\renewcommand{\baselinestretch}%
{#1}\small\normalsize} \spacingset{1}

\setcitestyle{authoryear,open={(},close={)}}
\date{} 
\begin{document}


\if0\blind
{
  \title{\bf A Bayesian Dirichlet Auto-Regressive Moving Average Model for Forecasting Lead Times}
\author[1,2]{Harrison Katz}
\author[2]{Kai Thomas Brusch}
\author[3]{Robert E. Weiss}

\affil[1]{Department of Statistics, UCLA}
\affil[2]{Data Science, Forecasting, \airbnbp}
\affil[3]{Department of Biostatistics, UCLA Fielding School of Public Health }
  \maketitle
} \fi

\if1\blind
{
  \bigskip
  \bigskip
  \bigskip
  \begin{center}
    {\LARGE\bf A Bayesian Dirichlet Auto-Regressive Moving Average Model for Forecasting Lead Times}
\end{center}
  \medskip
} \fi

\begin{abstract}

In the hospitality industry, lead time data is a form of compositional data that is crucial for business planning, resource allocation, and staffing. Hospitality businesses accrue fees daily, but recognition of these fees is often deferred. This paper presents a novel class of Bayesian time series models, the Bayesian Dirichlet Auto-Regressive Moving Average (B-DARMA) model, designed specifically for compositional time series. The model is motivated by the analysis of five years of daily fees data from Airbnb, with the aim of forecasting the proportion of future fees that will be recognized in 12 consecutive monthly intervals. Each day's compositional data is modeled as Dirichlet distributed, given the mean and a scale parameter. The mean is modeled using a Vector Auto-Regressive Moving Average process, which depends on previous compositional data, previous compositional parameters, and daily covariates. The B-DARMA model provides a robust solution for analyzing large compositional vectors and time series of varying lengths. It offers efficiency gains through the choice of priors, yields interpretable parameters for inference, and produces reasonable forecasts. The paper also explores the use of normal and horseshoe priors for the VAR and VMA coefficients, and for regression coefficients. The efficacy of the B-DARMA model is demonstrated through simulation studies and an analysis of Airbnb data.
\bigskip

\end{abstract}

{\renewcommand{\thefootnote}{}
\footnotetext{ 

}}

\newpage

\doublespacing

\section[Introduction]{Introduction}

Travel, entertainment, and hospitality businesses earn fees each day, however these fees cannot be recognized until later. The \textit{lead time} is the amount of time until fees earned on a given day can be recognized. Future dates that fees may be recognized are allocated into regular intervals forming a partition of the future, often weekly, monthly or annual time intervals. Each day, another vector of fee allocations is observed. The distribution of fees into future intervals can be analyzed separately from the total amount of fees; fractional allocations into future intervals sum to one and form a compositional time series. The basic compositional observation is a continuous vector of probabilities that a dollar earned today can be recognized in each future interval. We wish to understand the process generating the compositional time series and forecast the series on into the future. This information can be used by businesses for the allocation of resources, for business planning, and for staffing. 

We analyze five years of daily fees billed at \airbnb that will be recognized in the future. Future fees are allocated to one of 12 consecutive monthly intervals; fees beyond this range are small and ignored in this analysis. More granular inference on lead time of revenue recognition would not improve business planning or resource allocation though it substantially complicates modeling, model fitting, and communication of results. To forecast a current day’s allocations we have all prior days’ data and we have deterministic characteristics of days such as day of the week, season, and sequential day of the year.

A lead time compositional time series $\bfy_t = (y_{t1}, \ldots, y_{tJ})^\prime$ is a multivariate $J$-vector time series with observed data $y_{tj}$ where $t = 1, \dots, T$, indexes consecutive days or other time units, $j = 1, \dots, J$ indexes the $J$ future revenue recognition intervals, $0 < y_{tj} < 1$ and $\sum_{j=1}^J y_{tj} = 1$ for all $t$. A natural model for compositional data is the Dirichlet distribution which is in the exponential family of distributions and thus Dirichlet time series are special cases of generalized linear time series. \citet{benjamin2003generalized} proposed a univariate Generalized ARMA data model in a frequentist framework.

%
There are numerous books on Bayesian Analysis of Time Series Data 
    \citep{barber2011bayesian,berliner1996hierarchical,pole2018applied, koop2010bayesian, west1996bayesian, prado2010time} as well as papers on Bayesian vector auto-regressive (AR) (VAR) moving average (MA) (VMA) (ARMA)(VARMA)  
time series models \citep{spencer1993developing, uhlig1997bayesian, banbura2010large, karlsson2013forecasting}; and on 
Bayesian generalized linear time series models 
    \citep{brandt2012bayesian, 
roberts2002variational,
nariswari2019bayesian,
chen2016generalized,
mccabe2005bayesian,
berry2020bayesian,
nariswari2019bayesian,
fukumoto2019bayesian,
silveira2015bayesian,
west2013bayesian}. 

    
Dirichlet time series data are less commonly modeled in the literature.  \citet{grunwald1993time} proposed a Bayesian compositional state space model with data modeled as Dirichlet given a mean vector, with the current mean given the prior mean also modeled as Dirichlet. Grunwald does not use Markov chain Monte Carlo (MCMC) for model fitting. Similarly, \citet{da2011dynamic} proposed a state space Bayesian model for a time series of proportions, extended in \citet{da2015bayesian} to Dirichlets with a static scale parameter.  \citet{zheng2017dirichlet} propose a frequentist Dirichlet ARMA time series.  Much of the prior work on modeling compositional data and compositional time series transforms $\bfy_t$ from the original $J$-dimensional simplex to a Euclidean space where the data is now modeled as normal \citep{aitchison1982statistical, Cargnoni1997BayesianFO, ravishanker2001compositional, silva2001modelling, mills2010forecasting, barcelo2011compositional, koehler2010forecasting, kynvclova2015modeling, snyder2017forecasting, al2018compositional}. 

It would seem preferable to not transform the raw data before modeling and instead model the data $\bfy_t$ directly as Dirichlet distributed. Thus we propose a new class of Bayesian Dirichlet Auto-Regressive Moving Average models (B-DARMA) for compositional time series. We model the data as Dirichlet given the mean and scale, then transform the $J$-dimensional mean parameter vector to a $J-1$-dimensional vector. The distributional parameters are then modeled with vector auto-regressive moving average structure. We also model the Dirichlet scale parameter as a log linear function of time-varying predictors. 

We give a general framework, and present submodels motivated by our \airbnb lead time data. The B-DARMA model can be applied to data sets with large compositional vectors or few observations, offers efficiency gains through choice of priors and/or submodels,
and provides sensible forecasts on the \airbnb data.

We consider normal and horseshoe priors for the VAR and VMA coefficients, and for regression coefficients. Normal priors have long been a default for coefficients while the horseshoe is a newer choice which allows for varying amounts of shrinkage \citep{carvalho2009handling,carvalho2010horseshoe, huber2019adaptive, kastner2020sparse, banbura2010large} depending on the magnitude of the coefficients.

The next section presents the B-DARMA model. 
Section \ref{simulation} presents simulation studies comparing the B-DARMA to both a frequentist DARMA data model and a transformed data normal VARMA model. Section \ref{airbnb} presents analysis of the \airbnb data. The paper closes with a short discussion.

\section{A Bayesian Dirichlet Auto-Regressive Moving Average Model}

\label{bayesiandirichlet}

\subsection{Data model}

We observe a $J$-component multivariate compositional time series  $\bfy_t = (y_{t1}, \ldots, y_{tJ})^{\prime}$, observed at consecutive integer valued times $t = 1$ up to the most recent time $t = T$, where $0 < y_{tj} < 1$, $\bfone^{\prime} \bfy_t = 1$ where $\bfone$ is a $J-$vector of ones. We model $\bfy_t$ as Dirichlet with mean vector $\bfmu_t = (\mu_{t1}, \ldots, \mu_{tJ})^{\prime}$, with $0 < \mu_{tj} < 1$, $\bfone^{\prime}\bfmu_t = 1$, and scale parameter $\phi_t > 0$
\begin{align}
\bfy_t | \bfmu_t, \phi_t \sim \Dirichlet(\phi_t\bfmu_t), \label{ytmodel}
\end{align}
with density $f(\bfy_t|\bfmu_t, \phi_t) \propto \prod_{j=1}^J y_{tj}^{\phi_t\mu_{tj}-1}$. 

We model $\bfmu_t$ as a function of prior observations $\bfy_1, \ldots, \bfy_{t-1}$, prior means $\bfmu_1, \ldots, \bfmu_{t-1}$ and known covariates $\bfx_t$ in a generalized linear model framework. As $\bfmu_t$ and $\bfy$ are constrained, we model $\bfmu_t$ after reducing dimension using the \textit{additive log ratio} (alr) link 
\begin{align}
\bfeta_t \equiv \alr(\bfmu_t) = \left(\log\left(\frac{\mu_{t1}}{\mu_{tj^*}}\right),\ldots,\log\left(\frac{\mu_{tJ}}{\mu_{tj^*}}\right)\right) \label{mueta}
\end{align}
where $j^*$ is a chosen reference component $1 \le j^* \le J$, and the element of $\bfeta_t$ that would correspond to the $j^*$th element $\log (\mu_{j^*}/\mu_{j^*}) = 0$ is omitted. The \textit{linear predictor} $\bfeta_t$ is a $J-1$-vector taking values in $\mathbb{R}^{J-1}$. Given $\bfeta_t$, $\bfmu_t$ is defined by the inverse of equation \eqref{mueta} where $\mu_{tj} = \exp(\eta_{tj})/(1 + \sum_{j = 1}^{J-1} \exp(\eta_{tj}))$ for $j = 1, \dots, J$, $j \ne j^*$ and for $j = j*$, $\eta_{tj*} = 1/(1 + \sum_{j = 1}^{J-1} \exp(\eta_{tj}))$. 

We model $\bfeta_t$ as a Vector Auto-Regressive Moving Average (VARMA) process 
\begin{align}
    \bfeta_t =  \sum_{p=1}^P \pmb{A}_p ( \alr(\bfy_{t-p}) - \bfX_{t-p}\pmb{\beta} ) + \sum_{q=1}^Q \pmb{B}_q( \alr(\bfy_{t-q})- \bfeta_{t-q}) + \bfX_t \pmb{\beta} \label{eta_f}
\end{align}
for $t = m + 1, \dots, T$ where $m = \max(P,Q)$, $\pmb{A}_p$ and $\pmb{B}_q$ are $(J-1) \times (J-1)$ coefficient matrices of the respective Vector Auto-Regressive (VAR) and Vector Moving Average (VMA) terms, $\pmb{X}_t$ is a known $(J-1) \times r_\beta$ matrix of deterministic covariates including an intercept, and including seasonal variables and trend as needed, and $\bfbeta$ is an $r_\beta \times 1$ vector of regression coefficients. The form of an intercept in $\bfX_t$ is the $(J-1) \times (J-1)$ identity matrix $I_{J-1}$ as $J-1$ columns in $\bfX_t$. Given an $r_0 \times 1$ vector of covariates $\bfx_t$ for day $t$, the simplest form of $\bfX_t$ is $\bfX_t = I_{J-1} \otimes \bfx_t$ and $r_\beta = (J-1) * r_0$. The additive log ratio in \eqref{eta_f} is a multivariate logit link and leads to elements of matrices $\pmb{A}_p$ and $\pmb{B}_q$ and vector $\pmb{\beta}$ that are log odds ratios. Given $\bfeta_t$, $\bfmu_t$ gives the expected allocations of fees to components.

Scale parameter $\phi_t$ is modeled with log link as a function of an $r_\gamma$-vector of covariates $\bfz_t$, 
\begin{align}
    \phi_t = \exp(\bfz_t\bfgamma),
    \label{phi_f}
\end{align}
where $\bfgamma$ is an $r_\gamma$-vector of coefficients. In the situation of no covariates for $\phi_t$, $\log \phi_t = \gamma$ for all $t$.

Define the consecutive observations $\bfy_{a:b} = (\bfy_a, \dots, \bfy_b)^\prime$ for positive integers $a < b$. To be well defined, linear predictor \eqref{eta_f} requires having $m$ previous observations $\bfy_{(t-m):(t-1)}$, 
and corresponding linear predictors $\bfeta_{t-m}, \ldots, \bfeta_{t-1}$. In computing posteriors, we condition on the first $m$ observations $\bfy_{1:m}$ which then do not contribute to the likelihood. For the corresponding first $m$ linear predictors, on the right hand side of \eqref{eta_f}, we set $\bfeta_{1}, \ldots, \bfeta_{m}$ equal to $\alr(\bfy_{1}), \ldots, \alr(\bfy_{m})$ which effectively omits the VMA terms $\pmb{B}_l( \alr(\bfy_{t-l}))$ from \eqref{eta_f} when $t-l \le m$. In contrast, in \eqref{eta_f} the VAR terms and $\bfX_t\bfbeta$ are well defined for $t = 1, \ldots, m$.

Define the $C$-vector $\bftheta$ of all unknown parameters $\bftheta = (\bfA_{prs}, \bfB_{qrs}, \bfbeta^\prime, \bfgamma^\prime)^\prime$, where $r,s = 1, \ldots, J - 1$ index all elements of matrices $\bfA_p$ and $\bfB_q$, $p = 1, \ldots, P$, $q = 1, \ldots, Q$ and $C = (P+Q)*(J-1)^2 + r_\beta + r_\gamma$. Prior beliefs $p(\bftheta)$ about $\bftheta$ are updated by Bayes' theorem to give the posterior
\begin{align*}
    p(\bftheta |\bfy_{1:T}) = 
    \frac{p(\bftheta)p(\bfy_{(m+1):T}|\bftheta, \bfy_{1:m})}{p(\bfy_{(m+1):T} | \bfy_{1:m})},
\end{align*}
where $p(\bfy_{(m+1):T}|\bftheta, \bfy_{1:m}) = \prod_{t=m+1}^T p(\bfy_t | \bftheta, \bfy_{(t-m):(t-1)})$, $p(\bfy_t | \bftheta, \bfy_{(t-m):(t-1)})$ is the density of the Dirichlet in \eqref{ytmodel}, and the normalizing constant $p(\bfy_{(m+1):T}| \bfy_{1:m}) = \int p(\bftheta)p(\bfy_{(m+1):T}|\bftheta, \bfy_{1:m}) d\bftheta$. 
We wish to forecast the next $S$ observations $\bfy_{(T+1):(T+S)}$. These have joint predictive distribution 
\begin{align*}
    p(\bfy_{(T+1):(T+S)}| \bfy_{1:T}) = \int_{\bftheta} p(\bfy_{(T+1):(T+S)}|\bftheta)   p(\bftheta |\bfy_{1:T})d\bftheta. 
\end{align*}
The joint predictive distribution $p(\bfy_{(T+1):(T+S)}| \bfy_{1:T})$ can be summarized for example by the mean or median against time $t \in (T+1):(T+S)$ to communicate results to business managers.

Our data model can be viewed as a Bayesian multivariate Dirichlet extension of the Generalized ARMA model by \cite{benjamin2003generalized}. Similarly, \cite{zheng2017dirichlet} propose a DARMA data model whose link function in \eqref{mueta} does not have an analytical inverse, so needs to be approximated numerically. They view our data model as an approximation to theirs, noting that the resulting noise sequence from having the analytical inverse isn't a martingale difference sequence (MDS). This lack of an MDS complicates the investigation of the probabilistic properties of the series and the asymptotic behavior of their estimators.

From a Bayesian perspective, the need for an MDS for making inference is circumvented. Bayesian inference is based on the posterior distribution of the parameters, which combines the likelihood function (dependent on the data) and the prior distribution (representing our prior beliefs about the parameters). This allows us to make inferences regardless of whether the data form an MDS.

Furthermore, the concept of a ``noise sequence'' or residuals is somewhat different in the context of generalized linear models (GLMs) compared to location-scale models like linear regression. In GLMs, residuals are not explicitly defined, and the noise sequence discussed in the context of Zheng and Chen's model is an approximation. In our Bayesian approach, we do not need to define or consider the noise sequence at all, which simplifies the model and the analysis.
\subsection{Choice of link function}
The link function in \eqref{mueta} and \eqref{eta_f} can be replaced with other common simplex transformations such as the Centered Log-Ratio (CLR)

\begin{equation*}
\text{clr}(\mathbf{y}) = \left[ \ln\left(\frac{y_1}{g(\mathbf{y})}\right), \ln\left(\frac{y_2}{g(\mathbf{y})}\right), \ldots, \ln\left(\frac{y_J}{g(\mathbf{y})}\right) \right]
\end{equation*}
where 
\( g(\mathbf{y}) = \left( y_1 \times y_2 \times \ldots \times y_J \right)^{1/J} \).
Alternatively, the Isometric Log-Ratio (ILR) transformation can be used, with $j$-th component 
\begin{equation*}
y_j = \sqrt{\frac{r_j}{r_j + 1}} \log \left( \frac{g_j(\mathbf{y})}{\left( \prod_{i \in H_j} y_i \right)^{1/r_j}} \right)
\end{equation*}
where $g_j(\mathbf{y})$ is the geometric mean of a subset $S_j$ of $\mathbf{y}$, $H_j$ is the complement of $S_j$, and $r_j$ is the number of elements in $H_j$. The choice of subsets $S_j$ and $H_j$ can vary depending on the specific problem and interpretation requirements.
\cite{egozcue2003isometric} show that the coefficient matrices of the ALR, CLR, and ILR are linear transformations of each other. This means that the DARMA data models with these three link functions are equivalent, provided the same transformation is applied to the priors.

\subsection{Model selection}
For model selection, we use the approximate calculation of the leave-future-out (LFO) estimate of the expected log pointwise predictive density (ELPD) for each model which measures predictive performance \citep{bernardo2009bayesian,vehtari2012survey},
\begin{align*}
\elpd_{\LFO}= \sum_{t=L}^{T-M} \log p(\bfy_{(t+1):(t+M)}|\bfy_{1:t})
\end{align*}
where
\begin{align}
p(\bfy_{(t+1):(t+M)}|\bfy_{1:t})=\int_{\bftheta} p(\bfy_{(t+1):(t+M)}|\bfy_{1:t},\bftheta) p(\bftheta|\bfy_{1:t}) d\bftheta,
\label{elpdlfo}
\end{align}
where $M$ is the number of step ahead predictions and $L$ is a chosen minimum number of observations from the time series needed to make predictions. 
We use Monte-Carlo methods to approximate (\ref{elpdlfo}) with $S^*$ random draws from the posterior and estimate $p(\bfy_{(t+1):(t+M)}|\bfy_{1:t})$ as 
\begin{align*}
    p(\bfy_{(t+1):(t+M)}|\bfy_{1:t}) \approx \frac{1}{S^*}\sum_{s=1}^{S^*} p(\bfy_{(t+1):(t+M)}|\bfy_{1:t},\bftheta_{1:t}^{(s)})
\end{align*}
for $s=1,\dots,S^*$ draws from the posterior distribution $p(\bftheta|\bfy_{1:t})$. Calculating the ELPD LFO requires refitting the model for each $t \in (L,\dots, T-M)$, to get around this, we use approximate M-step ahead predictions using Pareto smoothed importance sampling \citep{ELPDlfo}.

\subsection{Priors}
A useful vague prior for the individual coefficients in $\bftheta$ is a proper independent normal $N(V_0,V)$ with varying $V_0$'s and $V$'s depending on prior beliefs about the coefficients. We take $V_0=.4,.1$ in our data analysis when we believe a priori that the coefficients will be positive and $V_0=0$ when we are unsure. For elements $a_{prs}$ and $b_{qrs}$ of $\bfA_p$ and $\bfB_q$ matrices, we take $V=.5^2$ as we expect those elements to be between $[-1,1]$. For elements of $\bfbeta$, we might for example let $V$ vary with the standard deviation of the covariate.

It may be thought that many elements of $\bftheta$ might be at or near zero. In this case, rather than a normal prior, we might consider a shrinkage prior like that proposed by \cite{carvalho2010horseshoe}, who propose a horseshoe prior of the form 
\begin{align*}
\theta_c| \tau, \lambda_{c^*} \sim N(0, \tau^2 \lambda_{c^*}^2) \\
\lambda_{c^*} \sim C^+(0,1).
\end{align*} for some $c^*=1,\dots, C^*$, $c=1,\dots,C$, and $C^* \leq C$. Each $\theta_c$ belongs to one of $c^*=1,\dots,C^*$ groups and each group gets its own shrinkage parameter $\lambda_{c^*}$ applied to it. We apply this with component specific shrinkage parameters in our data analysis.

\section{Simulation study}

\label{simulation}

In two simulation studies, we study the B-DARMA model's capacity to accurately retrieve true parameter values, a critical aspect for reliable forecasting. By comparing B-DARMA with established models on a number of metrics, we provide a robust assessment of both its estimation and forecasting potential. Each simulation study has $L = 400$ data sets indexed by $l = 1, \ldots, L$ with $T = 540$ observations $\bfy_{t}^{(l)}$, $t = 1, \ldots, T$ with $J = 3$ components, $j^* = 3$ as the reference component for all models, and $\bfY^{l} =(\bfy_1^{(l)}, \dots, \bfy_T^{(l)})$. We split the $540$ observations into a training set of the first $500$ observations and a test set of the last $40$ observations, the holdout data, which the model is not trained on and which is used to show the models' forecasting performance. 

In both studies, we compare B-DARMA to a non-Bayesian DARMA data model and to a non-Bayesian transformed-data normal VARMA (tVARMA) model that transforms $\mathbf{y}_{t}^{(l)}$ to $\alr(\mathbf{y}_{t}^{(l)}) \in \mathbb{R}^{J-1}$ and models $\alr(\mathbf{y}_t) \sim N_{J-1}(\bfeta_t, \mathbf{\Sigma})$, where $\Sigma$ is an unknown $J-1 \times J-1$ positive semi-definite matrix, and $\bfeta_t$ is defined in \eqref{eta_f}.
We use the VARMA function from the MTS package \citep{MTSPackage} in R \citep{Rbase} to fit the tVARMA models with maximum likelihood. For the non-Bayesian DARMA data model, the BFGS algorithm as implemented in optim in R is used for optimization. To compute the parameters’ standard errors, the negative inverse Hessian at the mode is calculated.

The data generating model (DGM) is a DARMA model in simulation 1 and a tVARMA model in simulation 2. We set $p = 1$, $q = 1$ for a DARMA$(1,1)$ (simulation 1) or tVARMA$(1,1)$ (simulation 2) data generating model. 

To keep the parameterization of the B-DARMA$(P,Q)$ model consistent with the parameterization of the MTS package, we remove $-\bfX_{t-p}\bfbeta$, $p = 1, \dots, P $ from the VAR term in \eqref{eta_f} and set $\bfeta_t = \sum_{p=1}^P \bfA_p ( \alr(\bfy_{t-p}) ) + \sum_{q=1}^Q \bfB_q( \alr(\bfy_{t-q})- \bfeta_{t-q}) + \bfX_t \bfbeta^*$, where $\bfbeta^* =  \bfbeta - \sum_{p=1}^P \bfA_p\bfbeta$ and we drop the asterisk on $\bfbeta$ for the remainder of this section.

All B-DARMA models are fit with STAN \citep{Rstan} in R. We run $4$ chains with $2000$ iterations each with a warm up of $1000$ iterations for a posterior sample of $4000$. Initial values are selected randomly from the interval $[-1,1]$. 

In all simulations, the covariate matrix $\bfX_t = \pmb{I}_2$, the $2\times 2$ identity matrix. Unknown coefficient matrices $\bfA$ and $\bfB$ have dimension $2\times 2$. The shared parameters of interest for the DARMA, B-DARMA, and tVARMA are $\bfbeta = (\beta_1, \beta_2)^\prime$, $\bfB = (b_{rs})$, and $\bfA_p = (a_{prs}), r, s = 1, 2$, $p=1$ and $\bfbeta = (\beta_1, \beta_2)^\prime$. We use the out of sample Forecast Root Mean Squared Error, $\FRMSE_j$, and  Forecast Mean Absolute Error, $\FMAE_j$ for each component $j$ as a measure of forecasting performance 
\begin{align*}
 \FRMSE_{j} = \left( \left ( \frac{1}{400} \right) \left( \frac{1}{40}\right)\sum_{l=1}^{400} \sum_{t = 501}^{540} (y_{tj}^{l} - \bar{\mu}_{tj}^{l})^2 \right)^{\frac{1}{2}} \\
  \FMAE_{j} = \left ( \frac{1}{400} \right) \left( \frac{1}{40}\right) \sum_{l=1}^{400} \sum_{t = 501}^{540} |y_{tj}^{l} - \bar{\mu}_{tj}^{l}|
 \end{align*}
where $\bar{\mu}_{tj}^{l}$ is the posterior mean of $\mu_{tj}$ or the maximum likelihood estimate in the $l$th data set. The tVARMA model has additional unknown covariance matrix $\pmb{\Sigma} = (\Sigma_{rs}), r, s = 1, 2$ a function of 3 unknown parameters, two standard deviations  $\sigma_1=\Sigma_{11}^{1/2} $, $\sigma_2=\Sigma_{22}^{1/2}$, and correlation $\rho = \Sigma_{12} / (\sigma_1 * \sigma_2)$, while the B-DARMA and DARMA have a single unknown scale parameter $\phi$. For the tVARMA data generating simulations we set $\sigma_1 = \sigma_2 = .05$, $\rho = .30$, and for DARMA generating models, we set $\phi = 1000$.

Priors in simulations 1 and 2 for the B-DARMA$(1,1)$ are independent $N(0,.5^2)$ for all coefficients in $\beta, A, B$, and a priori $\gamma = \log \phi \sim \gammadist(25/7,5/7)$ with mean $5$ and variance $7$.

For each parameter, generically $\theta$, with true value $\theta_{\text{true}}$ in simulation $l$, for Bayesian models we take the posterior mean $\bar{\theta}^{l}$ as the point estimate. For the 95\% Credible Interval (CI) $(\theta^{l}_{\text{low}}, \theta^{l}_{\text{upp}})$ we take the endpoints to be 2.5\% and 97.5\% quantiles of the posterior. For the tVARMA and DARMA models, we use the maximum likelihood estimates (MLE) as the parameter estimate, and 95\% confidence intervals (CI) calculated as maximum likelihood estimate plus or minus 1.96 standard errors.

For each parameter $\theta$ in turn, for the B-DARMA, tVARMA, and DARMA models, we assess bias, root mean squared error (RMSE), length of the $95\%$ CI (CIL), and the fraction of simulations \% IN where $\theta_{\text{true}}$ falls within the $95\%$ CI
 \begin{align*}
     \bias & = \frac{1}{400}\sum_{l=1}^{400} (\bar{\theta}^{l} - \theta_{\text{true}}) \\
     \RMSE & = \left(\frac{1}{400} \sum_{l=1}^{400} (\bar{\theta}^{l} -\theta_{\text{true}})^2\right)^{1/2} \\
     \CIL & =  \frac{1}{400} \sum_{l=1}^{400} \left(\theta^{l}_{\text{upp}} - \theta^{l}_{\text{low}} \right) \\
     \IN & = \frac{1}{400}\sum_{l=1}^{400} \bfone\{\theta^{l}_{\text{low}} \le \theta_{\text{true}} \le \theta^{l}_{\text{upp}}\} 
 \end{align*}
for the B-DARMA models and replace $\bar{\theta}^{l}$ with the MLE for the tVARMA and DARMA models. 
 
For all simulations, elements of $\bfbeta, \bfA, \bfB$ are set to be
 \begin{align*}
\beta_{1}  & = -0.07, \beta_{2} = 0.10, \\ 
a_{11} & = 0.95, a_{12} = -0.18, a_{21} = 0.3, a_{22} = 0.95, \\
b_{11} & = 0.65,  b_{12} = 0.15, b_{21} = 0.2, b_{22} = 0.65 .
\end{align*}

\subsection{Comparing B-DARMA, DARMA, and tVARMA Models in simulations 1-2}

Supplementary tables \ref{tab:dgp_DARMA} and \ref{tab:dgp_tvarma} provide summarized parameter recovery results, while table \ref{tab:prediction_sim} gives the $\FRMSE_j$ and $\FMAE_j$ for each component. When the data generating model is a DARMA(1,1), B-DARMA consistently outperforms tVARMA, with the B-DARMA(1,1) yielding an RMSE averaging at $40\%$ smaller, especially for the $\bfB$ matrix coefficients. The B-DARMA's performance aligns with that of the frequentist DARMA, each excelling in half of the ten considered coefficients. Both DARMA models showcase  coverage superior to the tVARMA.

For a tVARMA(1,1) data generating model, the tVARMA generally exhibits the smallest RMSE for all coefficients, barring $b_{11}$. The difference in RMSE is marginal, averaging at $2 \%$. The B-DARMA and frequentist DARMA compare similarly for $\bfbeta$ and $\bfA$, but the DARMA performs worse for $\bfB$. B-DARMA has better coverage than both tVARMA and frequentist DARMA for most coefficients excepting $\beta_2$, $b_{21}$, and $b_{22}$.

The out-of-sample prediction results are summarized in table \ref{tab:prediction_sim}. When the DGM is a DARMA(1,1), the B-DARMA outperforms both the frequentist DARMA model and the tVARMA in $\FRMSE$ and $\FMAE$ across all components, most notably for $y_3$. When the DGM is a tVARMA, for $y_1$ and $y_2$ the B-DARMA performs comparably to the tVARMA and outperforms the DARMA. For $y_3$, the B-DARMA significantly outperforms both the tVARMA and DARMA.

\section{\airbnb Lead Time Data Analysis}

\label{airbnb}

The \airbnb lead time data, $\bfy_t$, is a compositional time series for a specific single large market where each component is the proportion of fees booked on day $t$ that are to be recognized in 11 consecutive 30 day windows and 1 last consecutive 35 day window to cover 365 days of lead times.

\begin{figure}
    \includegraphics[scale=.75]{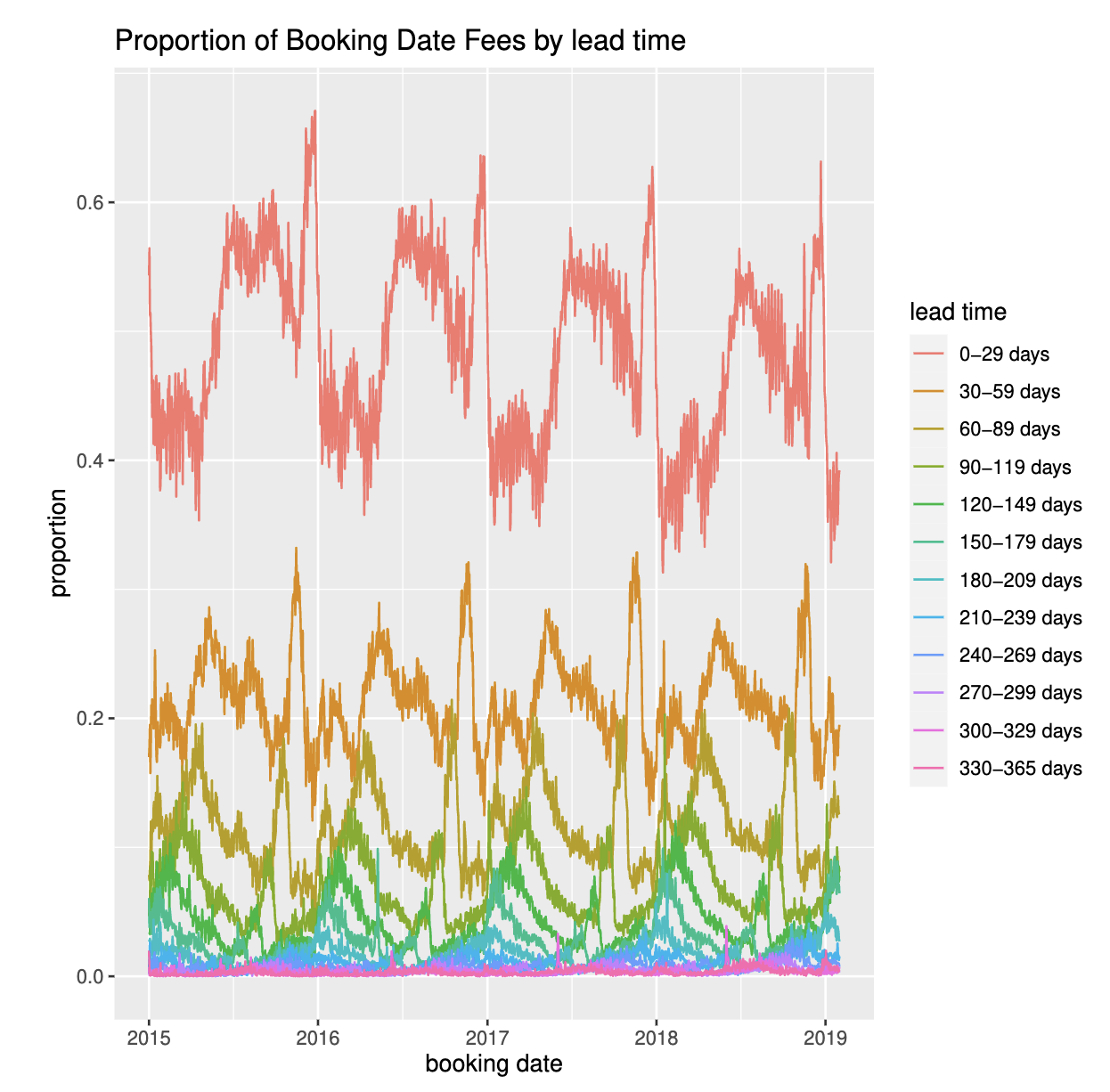}
    \caption{Airbnb data analysis - proportion of fees by lead time for a single large market from Jan 1, 2015 to Jan 31, 2019.}
    \label{fees}
\end{figure}

\begin{figure}
    \includegraphics[scale=.6]{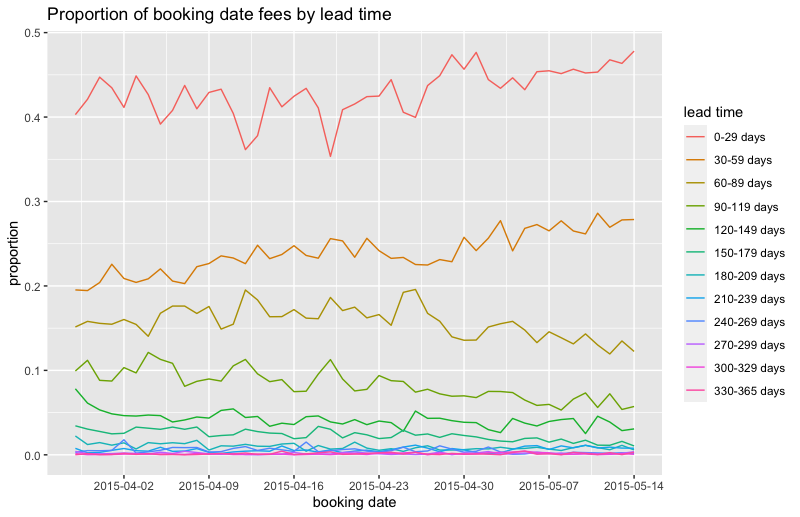}
    \caption{Airbnb data analysis- Proportion of fees for a single large market: weekly seasonal behavior.}
    \label{weekly_lead_time}
\end{figure}

Figures \ref{fees} and \ref{weekly_lead_time} plot the \airbnb lead time data from 01/01/2015 to 01/31/2019 and 03/26/2015 to 05/14/2015 respectively. 
There is a distinct repeated yearly shape in figure \ref{fees} and a clear weekly seasonality in figure \ref{weekly_lead_time}. Figure \ref{fees} shows a gradual increase/decrease in component sizes, most notably a decrease for the first window of $[0,29)$ days. The levels are driven by the attractiveness of certain travel periods with guests booking earlier and paying more for peak periods like summer and the December holiday season. The weekly variation shows a pronounced contrast between weekdays and weekends. Thus we want to include weekly and yearly seasonal variables and a trend variable in the predictors $\bfX_t$ in \eqref{eta_f}. We model each component with its own linear trend and use Fourier terms for our seasonal variables, pairs of $\left(\sin{\frac{2 k_{} \pi t}{w_{\season}}}, \cos{\frac{2 k_{} \pi t}{w_{\season}}}\right)$ for $k_{}=1,\dots,K_{\season} \leq \frac{w_{\season}}{2}$ where we take $w_{\season}=w_{\week} = 7$ for weekly seasonality and $w_{\season}=w_{\yearly} = 365.25$ for yearly seasonality. The orthogonality of the Fourier terms helps with convergence which is why we prefer it to other seasonal representations.

We train the model on data from  01/01/2015 to 1/31/2019, choose a forecast window of 365, and use 02/01/2019 to 01/31/2020 as the test set. All B-DARMA models are fit with STAN using the R interface where we run $4$ chains with $3000$ iterations each with a warm up of $1500$ iterations for a total of $6000$ posterior samples. Initial values are selected randomly from the interval $[-1,1]$.

\subsection{Model}

We use the approximate LFO ELPD of candidate models to first decide on the number of yearly Fourier terms, $K_{\yearly}$ given initial choices of $P$ and $Q$, then given the number of Fourier terms we decide on the orders $P$ and $Q$. We fix $K_{\season}=K_{\week}=3$, the pairs of Fourier terms for modeling weekly seasonality, for all models. We use an intercept and the same seasonal variables and linear trend for $\phi_t$. To define LFO ELPD, we take the step ahead predictions to be $M=1$, the minimum number of observations from the time series to make predictions to be $L=365$, and the threshold for the Pareto estimates to be $.7$ \citep{ELPDlfo,vehtari2015pareto} which resulted in needing to refit models at most $12$ times.

For all of the model selection process, we set independent $N(0,.5^2)$ priors for each $(a_{prs}),r,s=1,\dots,11 , p=1,2$ and $(b_{qrs}),r,s=1,\dots,11,q=1$, independent $N(0,1)$ priors on the Fourier coefficients, $N(0,.1)$ priors on the linear trend coefficients, and a $N(0,4)$ prior on the intercepts. We use these same priors for the seasonal, intercept and trend coefficients in $\gamma$ in \eqref{phi_f}. 

We fixed $P=1$, $Q=0$ and let $K_{\season}=K_{\yearly}$ take on increasing values starting with $3$ then sequentially $K_{\yearly} = 6,8,9,10$ with increasing values of LFO ELPD until 10 had worse LFO ELPD, so we stopped and took $K_{\yearly}=9$ (table \ref{tab:ELPD_airbnb}). We fixed $K_{\yearly}=9$ and similarly took $(P,Q)=(1,0)$ first then $(P,Q)= (1,1)$ and $(P,Q)=(2,0)$ with $(1,0)$ performing best (table \ref{tab:ELPD_airbnb}).  

We compare four different B-DAR(1) models plus a DAR(1) model and a tVAR(1) model. 
Model 1 (Horseshoe Full) has a horseshoe ($\tau=1$) prior on the coefficients in $\bfbeta$ and $\bfgamma$ with a separate shrinkage parameter $\lambda_{c^*}$ for the elements of $\bftheta$ that correspond to each $\eta_j$ or to $\phi$.  

As we expect the AR $a_{rs}$ elements to diminish in magnitude as the time difference $|r-s|$ increases, the other three models have varying prior mean and sd for $a_{rs}$ as functions of $|r-s|$. Model 2 (Normal Full) has an $a_{rr}\sim N(.4,.5^2)$ prior on its diagonal elements $r=1,\dots,11$, a $a_{r(r+1)},a_{r(r-1)}\sim N(.1,.5^2)$ prior on its two nearest neighbors, $r=1,\dots, 10 $ and $r=2,\dots,11$ respectively and a $N(0,.5^2)$ prior on the remaining elements of the $\bfA$ matrix.
 
 In contrast to models 1 and 2, where all parameters $a_{rs}$ in $\bfA$ are allowed to vary, models 3 and 4 fix some of the parameters $a_{rs}$ to $0$.  Model 3 (Normal Nearest Neighbor) only allows $a_{r(r+1)}, r=1,\dots, 10 ,a_{r(r-1)},r=2,\dots,11$ and $a_{rr},r=1,2,\dots,11$ to vary and model 4 (Normal Diagonal) only allows $a_{rr},r=1,2,\dots,11$ to vary. The remaining elements in model 3 $a_{r(r+k)}$ for $k \geq 2$ or $k \leq -2$ in the $\bfA$ matrix are set equal to 0. For model 4 all off diagonal elements, $a_{rs}$ $r\neq s$, are set to 0. The non-zero coefficients have the same priors as the Normal Full.

All 3 Normal models have the same priors on coefficients $\bfgamma$ and $\bfbeta$: $N(0,1)$ and $N(0,.1^2)$ for the coefficients of the Fourier and trend terms respectively. The scale model predictors include the same seasonal and trend terms and the corresponding priors on the coefficients are the same as for the $\bfeta$'s. The prior on the intercepts in $\bfbeta$ and $\bfgamma$ is $N(0,2^2)$.

Model 5 (DAR(1)) is the frequentist counterpart to our B-DAR(1) Normal Full model with the same seasonal and trend terms for $\eta_j$ and $\phi$. Model 6 (tVAR(1)) is a frequentist transformed VAR model which models $\alr(\mathbf{y}_t) \sim N_{J-1}(\bfeta_t, \mathbf{\Sigma})$. It has the same seasonal and trend variables in $\bfeta_t$.  Both models 5 and 6 are fit in R, model 5 with the BFGS algorithm as implemented in optim and model 6 with the VARX function in the MTS package.

\subsection{Results}

In the training set, the Horseshoe Full model has the largest LFO ELPD (supplementary table \ref{tab:ELPD}). The LFO ELPD for the Normal Full model and the Normal Nearest Neighbor model are close to the Horseshoe Full model with the Normal Diagonal model having a much worse LFO ELPD than the other models. 

For the test set, the Forecast Root Mean Squared Error ($\FRMSE_j$) and the Forecast Mean Absolute Error ($\FMAE_j$) for each component is in table \ref{tab:RSS_airbnb} as well as the total Forecast Root Mean Squared Error and Forecast Mean Absolute Error for each model. The Normal Full model performs best on the test set for most components and has the lowest total Forecast Root Mean Squared Error and smallest total Forecast Mean Absolute Error, with the Horseshoe Full model about $1.5\%$ and $.1\%$ worse respectively. The differences between the two full models (Model 1 and 2) and Models 3 and 4 are largest for the larger components as the $\FRMSE$'s for the smaller components are closer. For example, for the largest component $y_1$, the Normal Diagonal model performs over $7 \%$ worse than the Normal Full model. The Normal Nearest Neighbor model performs well considering it has $91$ fewer parameters than the Normal Full model, performing $3 \%$ worse in total than the Normal Full model.

When compared to the frequentist models, the Normal Full models exhibit superior performance in the largest components $y_1$ and $y_2$, having a smaller $\FMAE$ and a smaller $\FRMSE_j$. For the smaller components \(y_4\) through \(y_{12}\), the differences in $\FRMSE_j$ among the models are subtler, although the FMAE values continue to show a more pronounced difference across all components. The DAR(1) does perform better than the subsetted B-DAR(1) models with a total \(\FRMSE\) roughly \(2.5\%\) smaller than the Normal Nearest-Neighbor model and \(5 \%\) less than the Normal Diagonal model. The tVAR(1) performs significantly worse than all the DAR models, with a total \(\FRMSE\) about \(15\%\) larger than the worst performing DAR data model, suggesting the time invariant nature of \(\pmb{\Sigma}\) is inappropriate for the data.

\begin{figure}
    \includegraphics[scale=.6]{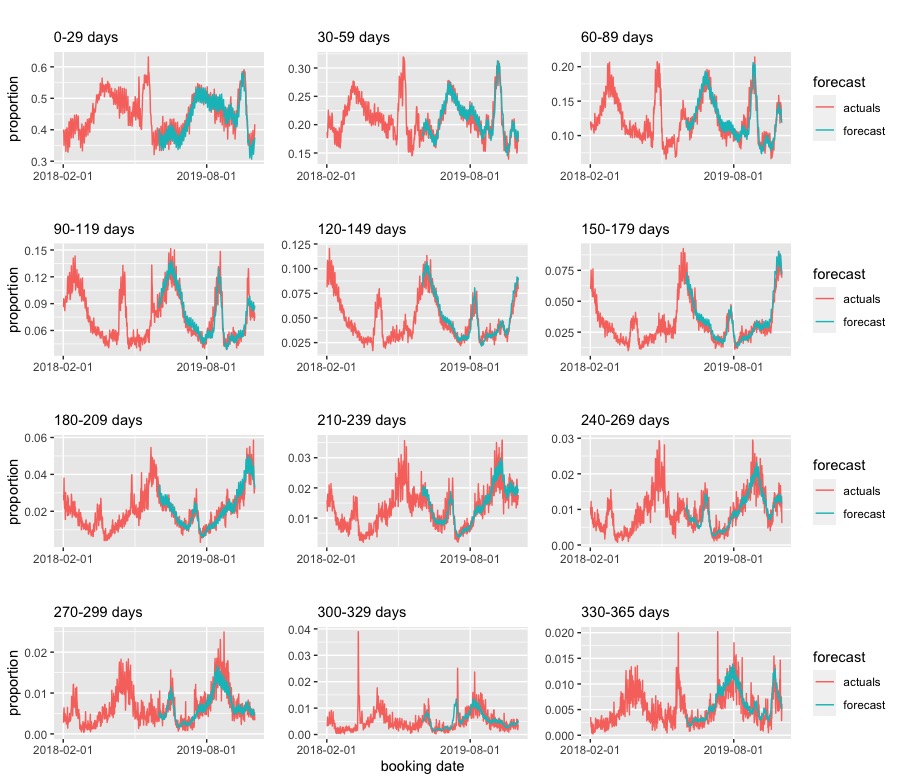}
        \caption{Airbnb data analysis -  Normal Full Model: one year of predictions (blue) from Feb 1, 2019 to Jan 31, 2020 and two  years of actuals (red) for each of the 12 components. }
        \label{prediction}
\end{figure}

\begin{figure}
\includegraphics[scale=.55]{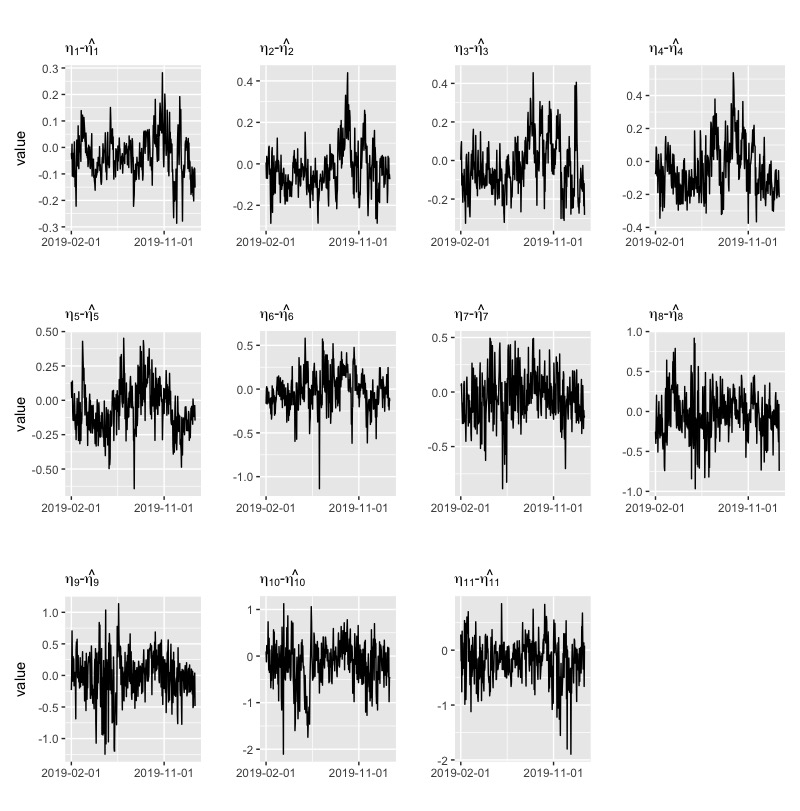}
    \caption{Airbnb data analysis -  Normal full model residuals on the additive log ratio scale, $\bfeta_t-\hat{\bfeta}_t$, for the test set, Feb 1, 2019 to Jan 31, 2020. }
    \label{residuals}
\end{figure}

Figures \ref{prediction} and \ref{residuals} plot out of sample forecasts ($\bar{\mu}_{tj}$) and residuals of $(\alr({\bfy_t})-\hat{\bfeta}_t)$ for the Normal Full model. 
The model captures most of the yearly seasonality and the residual terms exhibit no consistent positive or negative bias for any of the components as they are all centered near $0$.  Much of the remaining residual structure may be explained by market specific holidays which are not incorporated in the current model. 

The estimated yearly and weekly seasonality for each of the $\bfeta_{j}$'s (ALR scale) and $\phi$ (log scale) in the Normal Full model are detailed in supplementary figures \ref{eta_yearly} and \ref{eta_weekly}. Yearly variation is pronounced in larger components, with $\bfeta_1$ peaking in late December and troughing before the new year, while $\phi$ exhibits a simpler pattern with two high and one low period. Weekly seasonality displays a consistent weekday versus weekend behavior across all $\bfeta_j$ components, showing higher values on weekends. Supplementary table \ref{tab:summary_1} summarizes the posterior statistics for the intercepts and linear growth rates of $\bfeta_j$s and  $\log(\phi)$, showing a monotonic decrease in the intercepts and larger growth rates for smaller components. The growth rate of $\phi$ indicates less compositional variability over time. Lastly, the posterior densities for the elements $a_{rs}$ of the $\bfA$ matrix, shown in supplementary figure \ref{post}, reveal that larger components have pronounced coefficients on their own lag, $a_{rr}$, with diminishing magnitude for distant trip dates, while smaller components exhibit more uncertainty but generally maintain a strong positive coefficient on their own lag.

\section{Discussion}

\label{discussion}

There are distinct differences in the computational costs in fitting our DARMA model using frequentist or Bayesian methodologies, as well as the tVARMA model using the MTS package in R. Firstly, fitting the tVARMA model using the MTS package was fastest, attributable to its implementation of conditional maximum likelihood estimation with a multivariate Gaussian likelihood. This approach benefits from the mathematical simplicity of estimation under the normal model and its unconstrained optimization problem, resulting in significantly less computational overhead. This is in addition to the built-in performance optimizations found in specialized packages like MTS.

In contrast, the Bayesian DARMA model, implemented using Hamiltonian Monte Carlo (HMC), had higher computational costs. This is to be expected, given that HMC requires running many iterations to generate a representative sequence of samples from the posterior distribution. However, the B-DARMA fit faster than the frequentist DARMA. The frequentist DARMA model, fitted using the Broyden–Fletcher–Goldfarb–Shanno (BFGS) optimization method, was the most computationally demanding in our study. The complexity arises from the calculation of the Hessian matrix estimates, the computation of gamma functions in the Dirichlet likelihood, and the constraint of positive parameters. Notably, around 25\% of model fitting attempts in the simulation studies initially failed and required regenerating data and refitting the model, underscoring the computational challenges of this approach.

This paper presented a new class of Bayesian compositional time series models assuming a Dirichlet conditional distribution of the observations, with time varying Dirichlet parameters which are modeled with a VARMA structure. The B-DARMA outperforms the frequentist tVARMA and DARMA when the underlying data generating model is a DARMA and does comparably well to the tVARMA and outperforms the DARMA when tVARMA is the data generating model in the simulation studies. By choice of prior and model subsets, we can reduce the number of coefficients needing to be estimated and better handle data sets with fewer observations. This class of models effectively models the fee lead time behavior at \airbnbp, outperforming the DAR and tVAR with the same covariates, and provides an interpretable, flexible framework.  

Further development is possible. Common compositional time series have components with no ordering while the 12 lead time components in the \airbnb data set are time ordered. This suggests $\bftheta$ can be modeled hierarchically. The construction of the Nearest Neighbor and Diagonal models for $\bfA_1$ are examples of using time ordering in model specification. While the distributional parameters $\bfmu_t$ and $\phi_t$ are time-varying, the elements of $\bftheta$ are static and thus the seasonality is constant over time, which may not be appropriate for other data sets. We model $\phi_t$ with exogenous covariates, but our approach is flexible enough to allow for it to have a more complex AR structure itself. While not an issue for the \airbnb data set, the model's inability to handle 0's would need to be addressed for data sets with exact zeroes.

\section*{Acknowledgement}
The authors thank Sean Wilson, Jackson Wang, Peter Coles, Tom Fehring, Jenny Cheng, Erica Savage, and George Roumeliotis for helpful discussions. The authors would also like to thank the Editor, the Associate Editor, and three anonymous reviewers for their careful reading, comments, and suggestions to an earlier version of the paper. Special thanks are extended to Emily Lively, Lauren Mackevich, and Stephanie Chu for their indispensable support in navigating administrative, programmatic, and legal aspects, respectively, which significantly contributed to the timely completion and dissemination of this paper.

\section*{Code and Data Availability}
The primary dataset used in our study, ``A Bayesian Dirichlet Auto-Regressive Moving Average Model for Forecasting Lead Times,'' is not publicly available due to confidentiality constraints. However, the Stan code for the Bayesian Dirichlet Auto-Regressive Moving Average (B-DARMA) model is available for public access. It can be found at our GitHub repository: \href{https://github.com/harrisonekatz/B-DARMA-paper}{https://github.com/harrisonekatz/B-DARMA-paper}.

\bibliographystyle{chicago}
\bibliography{references}

\section*{Figures and tables}

\begin{table}[ht]
\small
\begin{tabular}{lrrrrrrrr}
  \hline
& DGM: & DARMA(1,1) & & & tVARMA(1,1) &  &     \\ 
& Model: & B-DARMA & DARMA & tVARMA & B-DARMA & DARMA & tVARMA  \\
  \hline
FRMSE &     $y_1$ &   0.1054 & 0.1091 &  0.1063 &     0.0566 &   0.0794 &    0.0566 \\
   &     $y_2$ &   0.1342 & 0.1367 &  0.1356 &     0.0748 &   0.0775 &    0.0748 \\
   &     $y_3$ &   0.1015 & 0.1082 &  0.1470 &     0.0548 &   0.0640 &    0.0787 \\
   \hline
  FMAE &      $y_1$ &   0.0770 & 0.0777 &  0.0800 &     0.0432 &   0.0633 &    0.0432 \\
    &      $y_2$ &   0.1004 & 0.1009 &  0.1020 &     0.0574 &   0.0652 &    0.0574 \\
    &      $y_3$ &   0.0766 & 0.0775 &  0.1102 &     0.0422 &   0.0529 &    0.0617 \\
   \hline 
 
\hline
\end{tabular}
\caption{\label{tab:prediction_sim} Simulation study Forecast Root Mean Squared Error (FRMSE) and Forecast Mean Absolute Error (FMAE) on the test set $(T=40)$. DGM is data generating model. Simulation study 1 is on the left and study 2 on the right.}
\end{table}

\begin{table}[ht]
\centering
\begin{tabular}{rrrrrr}
  \hline
(P,Q) & pairs of Fourier terms & ELPD diff & LFO ELPD \\
  \hline
(1,0) &9 &  0.0      & 70135.3   \\ 
(1,0) &10 & -23.3  & 70112.0    \\ 
(1,0) &8 & -83.2  & 70052.0 \\
(1,0) &6 & -594.7  & 69540.6     \\ 
(1,0) &3 & -1190.1   & 68945.2   \\
\hline
(1,1) & 9 & -101.7  &      70033.6 \\
(2,0) &9& -123.7 &  70011.5 \\
   \hline
\end{tabular}
\caption{Airbnb data analysis -  Leave-future-out expected log pointwise predictive density (LFO ELPD) and the LFO ELPD differences between the best performing B-DARMA model and candidate B-DARMA models. The first five candidate models fix $(P,Q)$ and vary pairs of Fourier terms for yearly seasonality and the last two candidate models fix the Fourier terms at 9 pairs and vary $(P,Q)$.}
\label{tab:ELPD_airbnb}
\end{table}

\begin{table}[ht]
\centering
\begin{tabular}{lrrrrrrr}
  \hline
  & & Normal & Horseshoe & Normal & Normal \\
  &  & Full &  Full &  Nearest-Neighbor &  Diagonal & DAR(1) & tVAR(1) \\ 
  \hline
      FRMSE &         \(y_1\)  & 0.0211 &    0.0213 &           0.0216 &   0.0225 &  0.0213 &   0.0301 \\
          &         \(y_2\)  & 0.0122 &    0.0124 &           0.0129 &   0.0132 &  0.0124 &   0.0145 \\
          &         \(y_3\)  & 0.0101 &    0.0101 &           0.0103 &   0.0106 &  0.0101 &   0.0118 \\
          &         \(y_4\)  & 0.0094 &    0.0094 &           0.0097 &   0.0099 &  0.0094 &   0.0110 \\
          &         \(y_5\)  & 0.0061 &    0.0061 &           0.0062 &   0.0063 &  0.0061 &   0.0076 \\
          &         \(y_6\)  & 0.0050 &    0.0050 &           0.0051 &   0.0052 &  0.0050 &   0.0056 \\
          &         \(y_7\)  & 0.0037 &    0.0037 &           0.0039 &   0.0040 &  0.0037 &   0.0042 \\
          &         \(y_8\)  & 0.0029 &    0.0029 &           0.0030 &   0.0031 &  0.0029 &   0.0033 \\
          &         \(y_9\)  & 0.0023 &    0.0023 &           0.0023 &   0.0023 &  0.0023 &   0.0026 \\
          &        \(y_{10}\)  & 0.0022 &    0.0022 &           0.0023 &   0.0023 &  0.0022 &   0.0023 \\
          &        \(y_{11}\)  & 0.0025 &    0.0023 &           0.0025 &   0.0025 &  0.0025 &   0.0025 \\
          &        \(y_{12}\)  & 0.0022 &    0.0022 &           0.0022 &   0.0022 &  0.0022 &   0.0030 \\
       \hline
       
          &      Total & 0.0796 &    0.0800 &           0.0821 &   0.0841 &  0.0801 &   0.0983 \\

    \hline \hline
FMAE & \(y_1\)     & 0.0170 & 0.0171 & 0.0174 & 0.0181 & 0.0172 &0.0247 \\  
 & \(y_2\)    & 0.0090 & 0.0094 & 0.0099 & 0.0111 & 0.0096 &0.0113 \\   
& \(y_3\)   & 0.0076 & 0.0079 & 0.0081 & 0.0083 & 0.0079 &0.0090  \\   
 &\(y_4\)  &   0.0067 & 0.0068 & 0.0070 & 0.0079& 0.0070 & 0.0085\\  
& \(y_5\)  &   0.0048 & 0.0049 & 0.0049 & 0.0051  & 0.0049 &0.0059 \\ 
 &\(y_6\)  &   0.0039 & 0.0039 & 0.0042 & 0.0042  & 0.0039 & 0.0044\\  
& \(y_7\)  &   0.0027 & 0.0027 & 0.0027 & 0.0030 & 0.0027 &0.0032 \\ 
 &\(y_8\)  &  0.0021 & 0.0021 & 0.0022 & 0.0023 & 0.0022 & 0.0025\\ 
& \(y_9\)  &   0.0018 & 0.0018 & 0.0018 & 0.0019 & 0.0018 &0.0019 \\ 
& \(y_{10}\)     &  0.0016 & 0.0016 & 0.0016 & 0.0016 & 0.0016 &0.0017 \\ 
& \(y_{11}\)     & 0.0017 & 0.0017 & 0.0017 & 0.0017 & 0.0017 & 0.0018\\  
 &\(y_{12}\)    & 0.0016 & 0.0016 & 0.0016 & 0.0017 &0.0016 & 0.0022 \\ 
 \hline
  & Total    & 0.0605 & 0.0615 & 0.0631 & 0.0669 & 0.0621 & 0.0771 \\

\end{tabular}
\caption{Airbnb data analysis - Forecast Root Mean Squared Error (FRMSE) and Forecast Mean Absolute Error (FMAE) for the 12 components and their total for the test set from Feb 1, 2019 to Jan 31, 2020.}
\label{tab:RSS_airbnb}
\end{table}

\newpage

\renewcommand{\thetable}{\arabic{table}}
\renewcommand{\tablename}{Supplementary Table}
\renewcommand{\figurename}{Supplementary Figure}

\begin{table}[ht]
\centering
\begin{tabular}{rrrrrrrrrrr}
  \hline
 par & model & true  & RMSE  & bias & coverage & length \\ 
  \hline
$\beta_1$ & B-DARMA & -.07 &   0.0083 & -0.0019 & 0.9450 & 0.0312 \\ 
& DARMA & & 0.0088  &-0.0015  &  0.9225 & 0.0312 \\
 & tVARMA & &  0.0112 & -0.0038 & 0.9000 & 0.0354 \\[1mm] 

$\beta_2$ & B-DARMA & .10 &    0.0081 & -0.0012 & 0.9425 & 0.0313 \\ 
& DARMA & & 0.0083  & -0.0007 &  0.9425 & 0.0312 \\
& tVARMA &  & 0.0096 & -0.0035 & 0.9425 & 0.0361\\[1mm]
 
 $a_{11}$ & B-DARMA & .95 &    0.0113 & -0.0021 & 0.9250 & 0.0385 \\
   & DARMA & &  0.0108 & -0.0025 & 0.9600 & 0.0395 \\ 
 & tVARMA &  & 0.0145 & -0.0036 & 0.8950 & 0.0428 \\[1mm]

$a_{12}$ & B-DARMA & -.18 &  0.0074 & -0.0014 & 0.9400 & 0.0287\\ 
  & DARMA & &  0.0079 & -0.0009 &0.9400 & 0.0294 \\ 
& tVARMA & &  0.0133 & -0.0013 & 0.8500 & 0.0316 \\

$a_{21}$ & B-DARMA & .30 &   0.0101 & -0.0020 & 0.9500 & 0.0373 \\ 
  & DARMA & & 0.0101  & -0.0009  &0.9500  & 0.0382 \\ 
& tVARMA &  & 0.0127 & -0.0041 & 0.9225 & 0.0436 \\

$a_{22}$ & B-DARMA & .95 & 0.0086 & -0.0021 & 0.9300 & 0.0311 \\ 
  & DARMA & & 0.0084  & -0.0018 & 0.9300 &0.0316  \\ 
& tVARMA &  & 0.0108 & -0.0028 & 0.9050 & 0.0323 \\

$b_{11}$ & B-DARMA & .65 & 0.0309 & -0.0064 & 0.9450 & 0.1244\\ 
  & DARMA & & 0.0308  & -0.0011 &  0.9500 &0.1286  \\ 
& tVARMA & &   0.1012 & -0.0759 & 0.5025 & 0.1593 \\

$b_{21}$ & B-DARMA & .20  & 0.0314 & 0.0003 & 0.9475 & 0.1168 \\ 
  & DARMA & & 0.0297  &  0.0018 &  0.9425 &0.1162  \\ 
& tVARMA & &  0.0907 & -0.0485 & 0.5650 & 0.1768 \\

$b_{12}$ & B-DARMA & .15 &  0.0286 & 0.0010 & 0.9525 & 0.1145 \\ 
  & DARMA & & 0.0289  & 0.0031 &0.9525  &  0.1139 \\ 
& tVARMA & & 0.0900 & -0.0514 & 0.5425 & 0.1645 \\ 

$b_{22}$ & B-DARMA & .65 &  0.0321 & -0.0040 & 0.9500 & 0.1283 \\ 
  & DARMA & & 0.0341  & 0.0013 & 0.9350 &  0.1272 \\ 
& tVARMA & & 0.1135 & -0.0868 & 0.4550 & 0.1575 \\

   \hline
\end{tabular}
\caption{\label{tab:dgp_DARMA}Simulation study 1 results (RMSE, bias, coverage and length of the $95$\% credible or confidence interval) for B-DARMA(1,1) tVARMA(1,1) and DARMA(1,1) when the data generating model is a DARMA(1,1) for the two regression coefficients in $\bfbeta$, the four $a_{rs}$ elements of the $A$ matrix, and four $b_{rs}$ elements of the $B$ matrix.}
\end{table}

\begin{table}[ht]
\centering
\begin{tabular}{rrrrrrrrrrr}
  \hline
 par & model & true  & RMSE &  bias  & coverage & length \\ 
  \hline
  $\beta_1$ & B-DARMA& -.07 &   0.0063 & -0.0012 & 0.9650 & 0.0254  \\ 
    & DARMA & &0.0065   & -0.0008  & 0.9275 & 0.0336  \\ 
  & tVARMA & &  0.0061 & -0.0013 & 0.9450 & 0.0238 \\ 

 $\beta_2$ & B-DARMA & .10 &  0.0069 & -0.0010 & 0.9625 & 0.0254   \\
   &  DARMA & &0.0065   & -0.0007 & 0.9400 &0.0334  \\ 
 & tVARMA &  & 0.0062 & -0.0010 & 0.9550 & 0.0241 \\

  $a_{11}$ & B-DARMA & .95 &  0.0125 & -0.0035 & 0.9525 & 0.0461  \\ 
    &  DARMA & & 0.0125  & -0.0038 & 0.9300& 0.0604 \\ 
  & tVARMA & &  0.0115 & -0.0022 & 0.9525 & 0.0431 \\

  $a_{12}$ & B-DARMA & -.18 &  0.0086 & -0.0004 & 0.9650 & 0.0344 \\
    &  DARMA & &  0.0091 & -0.0007 &  0.9225 &  0.0595 \\ 
  & tVARMA & &  0.0084 & -0.0007 & 0.9600 & 0.0321 \\

    $a_{21}$  & B-DARMA& .30 &  0.0115 & -0.0010 & 0.9600 & 0.0454 \\ 
      &  DARMA & &  0.0112 & -0.0009 & 0.9375& 0.0453 \\ 
  & tVARMA & &  0.0111 & -0.0007 & 0.9500 & 0.0437 \\ 
  $a_{22}$  & B-DARMA & .95 &   0.0090 & -0.0025 & 0.9475 & 0.0347 \\ 
    &  DARMA & &  0.0096 & -0.0040 &   0.9275 &0.0456  \\ 
  & tVARMA & &  0.0086 & -0.0016 & 0.9450 & 0.0325 \\ 

  $b_{11}$ & B-DARMA & .65 &  0.0342 & -0.0050 & 0.9475 & 0.1320  \\ 
    &  DARMA & & 0.0365  & -0.0015 &  0.9300    & 0.1728 \\ 
  & tVARMA & &  0.0348 & 0.0002 & 0.9350 & 0.1287 \\ 

  $b_{21}$ & B-DARMA & .20 & 0.0342 & -0.0027 & 0.9275 & 0.1318 \\ 
  &  DARMA & & 0.0357 & 0.0032 &  0.9150 &0.1718  \\ 

  & tVARMA & &  0.0337 & -0.0007 & 0.9475 & 0.1315 \\ 

  $b_{12}$ & B-DARMA & .15 &  0.0325 & -0.0011 & 0.9600 & 0.1320 \\ 
    &  DARMA & &  0.0346 &-0.0005 & 0.9125  & 0.1707 \\ 
  & tVARMA & & 0.0319 & 0.0012 & 0.9475 & 0.1278 \\ 

  $b_{22}$ & B-DARMA & .65 &   0.0354 & -0.0077 & 0.9375 & 0.1317 \\ 
    &  DARMA & & 0.0366   & 0.0003 & 0.8850  &0.1717  \\ 
  & tVARMA & & 0.0344 & -0.0008 & 0.9475 & 0.1318 \\ 

   \hline
\end{tabular}
\caption{\label{tab:dgp_tvarma}Simulation study 2 results (RMSE, bias, coverage and length of the $95$\% credible or confidence interval) for B-DARMA(1,1),  tVARMA(1,1) and DARMA(1,1) when the data generating model is a tVARMA(1,1) for the two regression coefficients in $\bfbeta$, the four $a_{rs}$ elements of the $A$ matrix, and four $b_{rs}$ elements of the $B$ matrix.}
\end{table}

\begin{table}[ht]
\centering
\begin{tabular}{lrrrrr}
  \hline
  Parameter & Model & Net Bias & RMSE Ratio & Net Coverage & Length Ratio \\ 
  \hline
  \(\beta_1\) & DARMA & -0.0004 & 1.06 & -0.0225 & 1.00 \\
  & tVARMA & -0.0019 & 1.35 & -0.0450 & 1.13 \\
  \(\beta_2\) & DARMA & 0.0005 & 1.02 & 0.0000 & 0.99 \\
  & tVARMA & -0.0023 & 1.18 & 0.0000 & 1.15 \\
  \(a_{11}\) & DARMA & -0.0004 & 0.96 & 0.0350 & 1.03 \\
  & tVARMA & -0.0015 & 1.28 & -0.0300 & 1.11 \\
  \(a_{12}\) & DARMA & 0.0005 & 1.07 & 0.0000 & 1.02 \\
  & tVARMA & 0.0001 & 1.80 & -0.0900 & 1.10 \\
  \(a_{21}\) & DARMA & 0.0011 & 1.00 & 0.0000 & 1.02 \\
  & tVARMA & -0.0021 & 1.26 & -0.0275 & 1.17 \\
  \(a_{22}\) & DARMA & 0.0003 & 0.98 & 0.0000 & 1.02 \\
  & tVARMA & -0.0007 & 1.26 & -0.0250 & 1.04 \\
  \(b_{11}\) & DARMA & 0.0053 & 1.00 & 0.0050 & 1.03 \\
  & tVARMA & -0.0695 & 3.27 & -0.4425 & 1.28 \\
  \(b_{21}\) & DARMA & 0.0006 & 0.95 & -0.0050 & 0.99 \\
  & tVARMA & -0.0482 & 2.89 & -0.3825 & 1.51 \\
  \(b_{12}\) & DARMA & 0.0001 & 1.01 & 0.0000 & 0.99 \\
  & tVARMA & -0.0504 & 3.15 & -0.4100 & 1.44 \\
  \(b_{22}\) & DARMA & 0.0073 & 1.06 & -0.0150 & 0.99 \\
  & tVARMA & -0.0828 & 3.54 & -0.4950 & 1.23 \\
  \hline
\end{tabular}
\caption{\label{tab:dgp_DARMA_metrics}Simulation study 1 results (Net Bias, RMSE Ratio, Net Coverage, and Credible/Confidence Interval Length Ratio) comparing B-DARMA(1,1) with tVARMA(1,1) and DARMA(1,1) when the data generating model is a DARMA(1,1) for the two regression coefficients in $\bfbeta$, the four $a_{rs}$ elements of the $A$ matrix, and four $b_{rs}$ elements of the $B$ matrix. RMSE and Length ratios greater than 1 and negative net coverage differences favor the B-DARMA model.}
\end{table}

\begin{table}[ht]
\centering
\begin{tabular}{lrrrrr}
  \hline
  Parameter & Model & Net Bias & RMSE Ratio & Net Coverage & Length Ratio \\ 
  \hline
  \(\beta_1\) & DARMA & -0.0004 & 1.03 & -0.0375 & 1.32 \\
  & tVARMA & -0.0001 & 0.97 & -0.0200 & 0.94 \\
  \(\beta_2\) & DARMA & 0.0003 & 0.94 & -0.0225 & 1.31 \\
  & tVARMA & 0.0000 & 0.90 & -0.0075 & 0.95 \\
  \(a_{11}\) & DARMA & -0.0003 & 1.00 & -0.0225 & 1.31 \\
  & tVARMA & 0.0013 & 0.92 & 0.0000 & 0.93 \\
  \(a_{12}\) & DARMA & -0.0003 & 1.06 & -0.0425 & 1.73 \\
  & tVARMA & -0.0003 & 0.98 & -0.0050 & 0.93 \\
  \(a_{21}\) & DARMA & 0.0001 & 0.97 & -0.0225 & 1.00 \\
  & tVARMA & 0.0003 & 0.97 & -0.0100 & 0.96 \\
  \(a_{22}\) & DARMA & -0.0015 & 1.07 & -0.0200 & 1.31 \\
  & tVARMA & 0.0009 & 0.96 & -0.0025 & 0.94 \\
  \(b_{11}\) & DARMA & 0.0035 & 1.07 & -0.0175 & 1.31 \\
  & tVARMA & 0.0052 & 1.02 & -0.0125 & 0.98 \\
  \(b_{21}\) & DARMA & 0.0059 & 1.04 & -0.0125 & 1.30 \\
  & tVARMA & 0.0020 & 0.98 & 0.0200 & 1.00 \\
  \(b_{12}\) & DARMA & 0.0006 & 1.06 & -0.0475 & 1.29 \\
  & tVARMA & 0.0023 & 0.98 & -0.0125 & 0.97 \\
  \(b_{22}\) & DARMA & 0.0080 & 1.03 & -0.0525 & 1.30 \\
  & tVARMA & 0.0069 & 0.97 & 0.0100 & 1.00 \\
  \hline
\end{tabular}
\caption{\label{tab:dgp_tvarma_metrics}Simulation study 2 results (Net Bias, RMSE Ratio, Net Coverage, and Length Ratio) comparing B-DARMA(1,1) with tVARMA(1,1) and DARMA(1,1) when the data generating model is a tVARMA(1,1) for the two regression coefficients in $\bfbeta$, the four $a_{rs}$ elements of the $A$ matrix, and four $b_{rs}$ elements of the $B$ matrix. RMSE and Length ratios greater than 1 and negative net coverage differences favor the B-DARMA model.}
\end{table}

\begin{table}[ht]
\centering
\begin{tabular}{rrrrr}
  \hline
 model & ELPD diff  &  LFO ELPD   \\
  \hline
Horseshoe Full &  0.0       & 70149.2  \\ 
Normal Full & -8.9 &  70140.3    \\ 
Normal Nearest-Neighbor & -37.3 & 70112.0   \\ 
Normal Diagonal & -186.2 & 69963.1    \\
   \hline
\end{tabular}
\caption{Airbnb data analysis -  Leave-future-out expected log pointwise predictive density (LFO ELPD) and the LFO ELPD differences between the best performing B-DARMA model and candidate B-DARMA models.}
\label{tab:ELPD}
\end{table}

\begin{table}[ht]
\centering
\begin{tabular}{rrrrrr}
  \hline
  & par & Mean & SD & Q2.5 & Q97.5 \\ 
  \hline
$\eta_1$ & intercept & -0.228 & 0.055 & -0.337 & -0.121 \\ 
$\eta_2$ &  & -0.526 & 0.071 & -0.666 & -0.389 \\ 
$\eta_3$ &  & -0.719 & 0.091 & -0.897 & -0.545 \\ 
$\eta_4$ &  & -1.033 & 0.112 & -1.254 & -0.811 \\ 
$\eta_5$ &  & -1.483 & 0.136 & -1.750 & -1.215 \\ 
$\eta_6$ &  & -2.409 & 0.169 & -2.740 & -2.070 \\ 
$\eta_7$ &  & -2.791 & 0.209 & -3.200 & -2.384 \\ 
$\eta_8$ &  & -2.819 & 0.249 & -3.299 & -2.330 \\ 
$\eta_9$ &  & -2.999 & 0.301 & -3.592 & -2.417 \\ 
$\eta_{10}$ &   & -3.064 & 0.349 & -3.743 & -2.380 \\ 
$\eta_{11}$ &  & -4.879 & 0.354 & -5.572 & -4.176 \\ 
$\phi$ &  & 6.746 & 0.023 & 6.701 & 6.791 \\ 
   $\eta_1$ & linear growth &  0.238 & 0.080 & 0.085 & 0.395 \\ 
   $\eta_2$ & &   0.607 & 0.100 & 0.411 & 0.804 \\ 
   $\eta_3$ & &   0.778 & 0.125 & 0.533 & 1.024 \\ 
   $\eta_4$ &   & 1.104 & 0.154 & 0.800 & 1.402\\ 
   $\eta_5$ &   & 1.674 & 0.189 & 1.306 & 2.050 \\ 
   $\eta_6$ &   & 2.235 & 0.239 & 1.766 & 2.702 \\ 
   $\eta_7$ & &  2.669 & 0.304 & 2.077 & 3.268 \\ 
   $\eta_8$ &   &  2.627 & 0.366 & 1.905 & 3.341 \\ 
   $\eta_9$ &  &   2.404 & 0.449 & 1.529 & 3.285 \\ 
   $\eta_{10}$ &  & 3.547 & 0.520 & 2.531 & 4.564\\ 
   $\eta_{11}$ &   & 7.297 & 0.512 & 6.295 & 8.314 \\ 
   $\phi$ &  & 6.824 & 0.268 & 6.297 & 7.347  \\

   \hline
\end{tabular}
\caption{Airbnb data analysis -  summary coefficients for the normal full model. Parameter (par), posterior mean (Mean), standard deviation (SD), and 95 \% CI. The linear growth rates are multiplied by $10^{4}$.}
\label{tab:summary_1}
\end{table}

\begin{figure}
    \includegraphics[scale=.55]{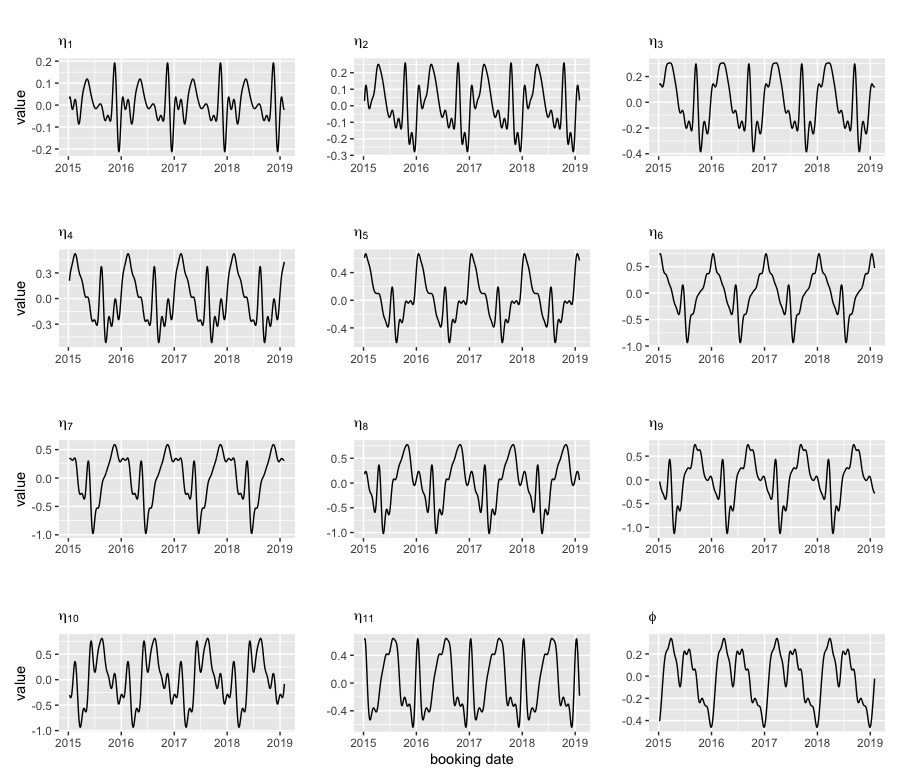}
        \caption{Airbnb data analysis - plot of the posterior mean yearly seasonal variation of the 11 components on the ALR scale and of the yearly seasonal variation for $\phi$ on the log scale for the normal full model Jan 1, 2015 to Jan 31, 2019.}
        \label{eta_yearly}
\end{figure}

\begin{figure}
    \includegraphics[scale=.55]{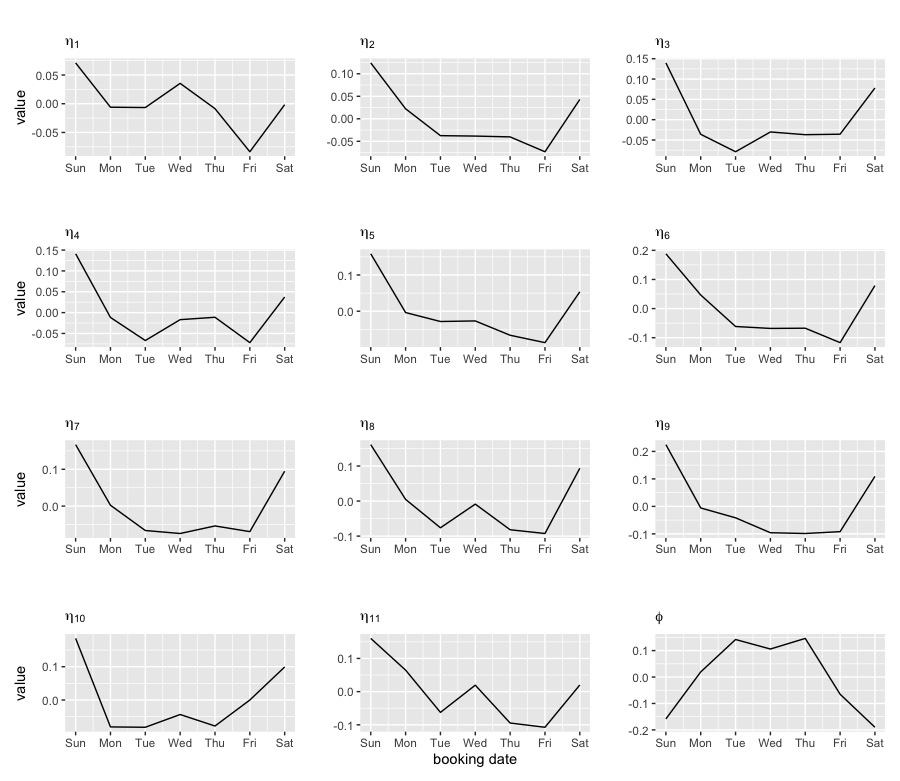}
        \caption{Airbnb data analysis - plot of the posterior mean weekly seasonality ($\eta$ ALR scale, $\phi$ log scale) for the normal full model.}
        \label{eta_weekly}
\end{figure}

\begin{figure}
    \centering
    \includegraphics[scale=.6]{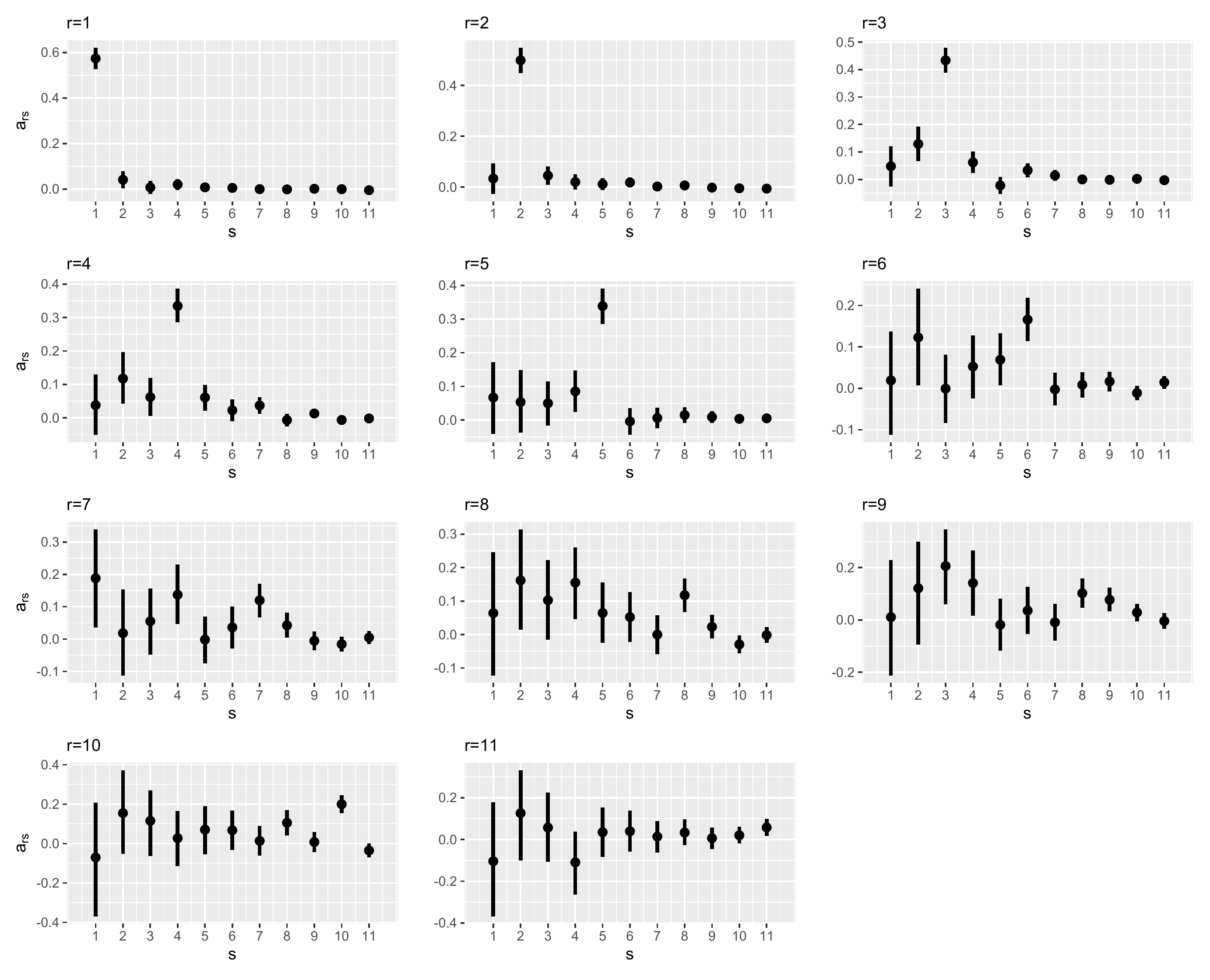}
    \caption{Airbnb data analysis - posterior density of elements $a_{rs}$ in VAR $A_1$ matrix. The center point of each density plot is the median of the posterior distribution and line segments are 95\% credible intervals. }
    \label{post}
\end{figure}

\end{document}